# General Doppler Shift Equation and the Possibility of Systematic Error in Calculation of Z for High Redshift Type Ia Supernovae

Steven M Taylor
smtygc@umsl.edu

**Abstract**

Systematic error in calculation of z for high redshift type Ia supernovae could help explain unexpected luminosity values that indicate an accelerating rate of expansion of the universe.

**Introduction**

The general form of the relativistic Doppler shift equation is

$$\nu' = \nu_0 \gamma (1 - \beta \cos\theta), \qquad (1)$$

where $\gamma = \dfrac{1}{\sqrt{1-\beta^2}}$ and $\beta = \dfrac{u}{c}$ with u being velocity of source.

With an emission angle $\theta = 0°$ the general form reduces to the familiar

$$\nu' = \nu_0 \frac{(1-\beta)^{\frac{1}{2}}}{(1+\beta)^{\frac{1}{2}}}. \qquad (2)$$

The condition $\theta = 0°$ corresponds to an emission antiparallel to the source's velocity vector and is typically assumed for astronomical purposes.

With the assumption $\theta = 0°$ redshift parameter is defined as

$$z = \frac{(1+\beta)^{\frac{1}{2}}}{(1-\beta)^{\frac{1}{2}}} - 1. \qquad (3)$$

**Evidence of Accelerating Universe and Possible Systematic Error**

The primary evidence of an accelerating rate of expansion of the Universe is that measurements of apparent magnitude of some high-z, type Ia supernovae are fainter than would be expected for non-accelerating cosmological models. [1]

Perlmutter and Schmidt of the Cosmology Supernovae Project have noted that along with other possible sources of systematic errors, gravitational lensing may contribute to a change in luminosity of high-redshift supernovae. Citing several authors, they note that as radiation traverses the large scale structure from where it is emitted and where it is detected, it could be lensed as it encounters fluctuations in gravitational potential. Some images could be demagnified as their light passes through under-dense regions. It is also noted that it would also be possible for a light path to encounter denser regions magnify the image. It is noted that such an effect may limit the accuracy of luminosity distance measurements.[2]

A lower luminosity in relation to z is to date the strongest evidence of an accelerated expansion rate for the Universe. In the same sense that a change in luminosity due to reasons other than distance, such as gravitational lensing could produce systematic error, so could a false z measurement.

Lowered luminosity would be consistent with a false z measurement if that measurement was less redshifted due to reasons extraneous to the expansion rate of the universe as presented by cosmological models.

Whether by gravitation or other effect, any canting in angle of emission of light from a receding source will cause an increase in frequency as seen by an observer. Taking the derivative of the general relativistic Doppler shift equation with respect to $\theta$ yields:

$$\frac{\partial v'}{\partial \theta} = v_0 \gamma \beta \sin \theta. \tag{4}$$

Since $\theta = 0°$ is an absolute minimum, any deviation from that angle results in a higher $v'$.

**Example**

According to Perivolaropoulos, the Gold supernova data set of 157 points show that transition from a decelerating towards and accelerating universe to be at z= $0.46 \pm 0.13$. Using a graph of apparent magnitude vs. redshift based on the Gold data, a supernova with an approximate 44 apparent magnitude and measured redshift parameter of $z \cong 0.95$, lies on the curve for an accelerating universe. If it did have a redshift parameter of $z \cong 1.30$, the supernova would lie on curve consistent with a decelerating universe.[3]

Using equation (3) a measured parameter of $z \cong 0.95$ corresponds to $\beta = 0.58$, and likewise a parameter of $z \cong 1.30$ corresponds to $\beta = 0.68$.

Assuming, for the sake of argument, that the Universe is decelerating and the supernova does have a parameter of $z \cong 1.30$, we can calculate the canting in angle of emission from $\theta = 0°$ that allows an observer to measure a parameter of $z \cong 0.95$.

Using $\nu_0 = 1$ Hz to simplify, and $\beta = 0.68$ (corresponding to z =1.30), equation 2 yields $\nu' = 0.4364$ Hz.

Likewise with $\nu_0 = 1$ Hz to simplify, and $\beta = 0.58$ (corresponding to z =0.95), equation 2 yields $\nu' = 0.5156$ Hz.

$$\Delta \nu' = 0.0792 \text{ Hz} \tag{5}$$

Inserting $\beta = 0.68$ into equation 4, yields

$$\frac{\partial \nu'}{\partial \theta} = \frac{0.68 \sin \theta}{(1 - 0.68^2)^{\frac{1}{2}}} . \tag{6}$$

Using 0.0792 Hz for $\dfrac{\partial \nu'}{\partial \theta}$ (5) into equation 6 yields an emission angle of $\theta \cong 4.90°$.

## Conclusion

Given the large distances that light from high z supernovae travel, and the modest canting from $\theta = 0°$ in emission angle required to help explain decreased luminosity for high redshift supernovae, the possibility of systematic error in z measurement for high redshift supernova should be further investigated.